\documentclass[article]{elsarticle}

\usepackage{eurosym}
\usepackage{newlfont}
\usepackage{graphicx}
\usepackage{amsmath}
\usepackage[latin1]{inputenc}
\usepackage{float}
\usepackage{color}

\journal{Physica A}

\begin{document}

\begin{frontmatter}

\title{Jarzynski equality for superconducting optical cavities: an alternative path to determine Helmholtz free energy}
\author{Josiane Oliveira Rezende de Paula}
\address{Escola Estadual C\^onego Luiz Vieira da Silva, Ouro Braco, MG 36420-000, Brazil }

\author{ J. G. Peixoto de Faria}
\address{Departamento de Matem\'atica, Centro Federal de Educa\c{c}\~ao Tecnol\'ogica de Minas Gerais, Belo Horizonte, MG, 30510-000, Brazil }

\author{J. G. G. de Oliveira Jr.}
\address{Departamento de Ci\^encias Exatas e Tecnol\'ogicas,
Universidade Estadual de Santa Cruz, 45.662--900, Ilh\'eus -- BA --
Brazil}

\author{Ricardo de Carvalho Falc\~ao}
\address{Departamento de Estat\'istica, F\'isica e Matem\'atica, Universidade
Federal de S\~ao Jo\~ao Del Rei, C.P. 131,Ouro Branco, MG, 36420-000,
Brazil}

\author{Ad\'elcio C. Oliveira}
\address{Departamento de Estat\'istica, F\'isica e Matem\'atica, Universidade Federal de S\~ao Jo\~ao Del Rei, C.P. 131,Ouro Branco, MG, 36420-000,
Brazil }

\begin{abstract}
A superconducting cavity model was proposed as a way to experimentally investigate the work performed in a quantum system. We  found a simple mathematical relation between the free energy variation and visibility measurement in quantum cavity context. If we consider the difference of Hamiltonian at time $t_0$ and $t_\lambda$ (protocol time) as a quantum work, then the Jarzynski equality is valid and the visibility can be used to determine the work done on the cavity.
\end{abstract}


\end{frontmatter}




\setcounter{MaxMatrixCols}{10}

\section{Introduction}

Fluctuation theorems have been developed to describe systems far
from equilibrium, that is the case of Jarzynski equality (JE)
\cite{jarzy,JAR1,JAR2,JAR3} and Crooks relation \cite{Crooks} and Bochkov-Kuzovlev \cite{Bochkov}. The classical JE is a relation between the free energy difference of two equilibrium states ($\Delta F$) and the work ($W$) averaged over all possible paths of a nonequilibrium process linking them. Mathematically the JE is
\begin{equation}
e^{-\beta\Delta F} = \langle e^{-\beta W} \rangle.
\end{equation}

The JE was developed assuming that the system is isolated from the reservoir while the protocol is performed. Morgado and Pinto \cite{Morgado} have obtained JE for a massive Brownian particle connected to internal and external springs, their result does not depend on the decoupling of system and bath along the protocol time, a Brownian particle was also experimentally investigated in context of JE \cite{EXJAR1}.  Minh and  Adib \cite{Minh} have used path integral formalism and demonstrated that the validity of JE  in the context of Brownian particle subject a class of harmonic potential.

Experimentally some important results were achieved,  Liphardt and collaborators \cite{Liphardt} demonstrated the validity of JE by mechanically stretching a single molecule of RNA reversibly and irreversibly between two conformations. Toyabe and collaborators \cite{JAR1} have investigated experimentally the JE for a dimeric particle comprising polystyrene beads by attaching it to a glass surface of a chamber filled with a buffer solution, they have found a discrepancy  smaller than $3\%$ between the observed result and what was expected with JE. Douarche and collaborators \cite{EXJAR3} have experimentally checked the Jarzynski equality and the Crooks \cite{Crooks} relation on the thermal fluctuations of a macroscopic mechanical oscillator in contact with a heat reservoir and found a good agreement with JE and crooks relation. Hoang et al. \cite{Hoang} have performed an experimental test of JE and  Hummer-Szabo  relation \cite{Hummer} using an optically levitated nanosphere. These, among many others experimental investigation consolidates the JE in the classical domain

In this work, we use the quantum analog of Jarzynski equality (JE) to propose a way to obtain experimentally the work performed in a quantum system. The Quantum version of (JE) is a controversy area, the first attempts to derive Jarzynski equalities for quantum systems failed \cite{Bochkov,Allahverdyan,Yukawa} leading to misleadingly believed that the equality was not valid for quantum systems. In some of this earlier derivations of Jarzynski equation for quantum systems a work operator was defined  \cite{Bochkov,Yukawa,Allahverdyan,ENGEL, Gelin}, but this work definition is not, in general, a quantum observable \cite{Talkner} this is due to the fact that work characterize a process rather than an instantaneous state of the system. This earlier attempts had led to quantum corrections to the classical Jarzynski result and the classical result was recovered only when the Hamiltonian in a time $t$ commutes with itself in a time $t'$ \cite{ENGEL}.

Recently the discussion has been changed to how to define operational ways of measuring work since Jarzynski's equality has already been obtained for closed quantum systems \cite{Kurchan,Tasaki,Talkner,Talkner1,Talkner4} for open systems \cite{Esposito, Crooks, Talkner5} even for systems with strong couplings \cite{Campisi}. Most of these proposals are linked to the question of measuring energy in two moments, which from a quantum point of view introduce several questions since a quantum systems have a dynamical behavior that is affected by the measurements, thus since one performs energy measurements the system state changes, this problem is circumvented if one use non-demolition measurements, in reference \cite{Paz} they show that POVM (positive operator valued measure) can be used to sample the work probability distribution. Experimentally, some advances have been achieved, An and collaborators \cite{EXJAR2} have investigated experimentally the JE in the quantum domain.  They have used $^{171}Yb_+$ ion trapped in a harmonic potential and perform projective measurements to obtain phonon distributions of the initial thermal state, they have concluded that JE still valid, a similar result was obtained by measuring a single-molecule \cite{EXJAR4}.

In this work, we study a transition between two equilibrium states
of a quantum system, namely a quantum harmonic oscillator coupled to
a thermal bath. This model can be implemented with a   cavity quantum
electrodynamics (CQED) \cite{cavidade}. The protocol can be executed by injecting a coherent field in the cavity. The CQED experimental setup
was widely  used to explore quantum mechanical foundations with many
interesting results (see \cite{Nogues,
Davidovich2016,Haroche1991,rydberg} and references therein). Even
for a more realistic model \cite{Faria1999}, that consider
environment action, the quantum nature of the electromagnetic field was demonstrated. Experimentally, the initial state was prepared in a pure state \cite{Haroche1991}, usually in a vacuum. We consider a thermal state, as the initial state, and the work is given by the difference of cavity's Hamiltonian $\Delta H = H(\tau)-H(0)$ where $\tau$ is a time bigger than the protocol time.
The thermal state is not a guarantee of a ``classical state'' \cite{Lemos2018}, but surprisingly, for the work as defined above, the JE is valid in all quantum domain \cite{Paz,Talkner,Talkner2,Talkner3}. Cerisola and collaborators \cite{Cerisola} have shown that JE is valid in quantum domain for a more general measurement class named ``a quantum work meter''.  Assuming that JE is valid, we show that the free energy variation and also the mean $\langle e^{-\beta W} \rangle$ can be simply inferred by a measurement of fringes visibility in the context of CQED.

\section{Quantum Jarzynski Equality}
\label{QJE}

We consider a Quantum analog of the model studied by \cite{Hijar}. It consists of $N$ non-interacting harmonic oscillators all initially in thermal e\-qui\-li\-brium at temperature $T$, then the partition function is
\begin{equation}
Z(0)=\prod_{n=1}^{N}Z_n(0)
\end{equation}
with
\begin{equation}
Z_n(0)=\frac{\exp \left( -\frac{1}{2}\beta \hbar \omega_{0}
    \right) }{1-\exp \left( -\beta \hbar \omega_{0} \right) }.
\end{equation}
After the action of the protocol, the equivalent quantum
Hamiltonian of $n$th oscillator for $t^{\prime }<t$ is
\begin{equation}
\hat{H}_{n}(t^{\prime })=\frac{\hat{p}_{n}^{2}}{2m}+\frac{1}{2}m\omega_{0} ^{2}\hat{x%
}_{n}^{2}+l\hat{x}_{n}L\left( t^{\prime }\right) .  \label{HAMLC}
\end{equation}
%
Then its eigenvectors are the same of harmonic oscillator and
the energies are
\begin{equation}
E_{n}(t^{\prime})=\left( j +\frac{1}{2}\right) \hbar \omega_{0} -\frac{1}{2}\frac{%
l^{2}L^{2}(t^{\prime })}{m\omega_{0} ^{2}},
\end{equation}
where $j$ are positive integers. Thus the partition function reads
\begin{equation}
Z(t^{\prime})=\prod_{n=1}^{N}Z_n(t^{\prime}),
\end{equation}
with
\begin{equation*}
Z_{n}(t^{\prime})=\exp\left[\frac{\beta l^{2}L^{2}(t^{\prime})}{2m\omega_{0} ^{2}}\right]
\left[ \frac{\exp\left(-\frac{1}{2}\beta\hbar\omega_{0}\right)}{
1-\exp\left(-\beta\hbar\omega_{0}\right)}\right] ,
\end{equation*}
and Helmholtz free energy to the $n$-th oscillator of the system is
\begin{equation}
F_{n}= \dfrac{\hbar\omega_0}{2}-\dfrac{l^2L^2(t^{\prime})}{2m\omega_0}
+
\dfrac{1}{\beta}\ln \left[ 1-\exp \left( -\beta \hbar \omega_{0} \right)
\right] .
\end{equation}
Again, the protocol changes $L$ parameter from $L_{0}$ to $L_{1}$. Thus,
\begin{equation}
\Delta F_{n}=\frac{l^{2}}{2m\omega_{0} ^{2}}\left(
L_{0}^{2}-L_{1}^{2}\right) . \label{DIFLC}
\end{equation}
It is easy to see that the variation of the Helmholtz free energy to the system will be
\begin{equation}
\Delta F=\sum_{n=1}^{N} \Delta F_{n} .
\end{equation}
Since we are dealing with $N$ non-interacting harmonic oscillators, without loss of generality, we can restrict our analysis to a single oscillator of system. We will do this from now on.

\subsection{Quantum work}



The Hamiltonian ($\hat{H}(t^{\prime})$) to a single oscillator of system is
\begin{equation}
\hat{H}_{n}(t^{\prime})=\frac{\hat{p_{n}}^2}{2}+\frac{\omega_{0} ^{2}}{2}\left[ \hat{x_{n}}%
^{2}+\frac{2l\hat{x_{n}}L(t^{\prime })}{\omega_{0} ^{2}}\right].
\label{HAMOP}
\end{equation}
where we set $m=\hbar=1$. We can also write the Hamiltonian in terms of creation and annihilation operators, and it will be useful in the next sections, it is given by

\begin{equation}
\hat{H}_{n}(t^{\prime})=\omega_{0} \left(\hat{a_{n}}^\dag \hat{a_{n}}+\frac{1}{2}\right)+\tilde{L}(t^{\prime})\left(\hat{a}+\hat{a}^\dag\right).
\label{HAMOP22}
\end{equation}
where  $\tilde{L}(t^{\prime})=\sqrt{\frac{\hbar}{2 m\omega_{0}}}L(t^{\prime})$.

We assume that the system environment coupling is not relevant during the
protocol time, if we assume that the necessary work \cite{Hijar} to change $%
L_{0}\longmapsto L_{1}$ is the same as the system energy variation $\Delta E$,
then at $t=0$ we find an energy $E_{n}^{(0)}$ with a probability given by
\begin{equation}
P_{n}^{(0)}=\frac{\exp \left( -\beta E_{n}^{(0)}\right) }{Z}\text{ .}
\end{equation}
After a time $t_{f}$ the system is in a state $\hat{U}%
(t_{f})\left| \psi _{n}^{0}\right\rangle$ and the transition probability is
\begin{equation}
w_{mn}=\left\vert \left\langle\psi _{m}^{(f)}\right| \hat{U}(t_{f})\left| \psi
_{n}^{(0)}\right\rangle\right\vert ^{2}.
\end{equation}
Here, $\hat{U}(t_{f})$ is
\begin{equation}
\hat{U_{n}}(t_{f})=T_{>}\exp \left[ -\frac{i}{\hbar }\int_{0}^{t_{f}}dt^{\prime }%
\hat{H}_{n}(t^{\prime })\right] ,
\end{equation}
where $T_{>}$ denotes time ordering operator. Finally, we obtain
\begin{equation}
\left\langle \exp (-\beta \Delta E)\right\rangle =\underset{n}{\sum }%
P_{n}^{(0)}\underset{m}{\sum }w_{mn}\exp (-\beta \Delta E)
\end{equation}
with $\Delta E = E_m^{\left(f\right)} - E_n^{\left(0\right)}$. After some manipulations \cite{Hijar} we get
\begin{equation}
\left\langle \exp (-\beta \Delta E)\right\rangle =\exp \left[
\frac{N \beta l^{2}}{2\omega_{0} ^{2}}\left(
L_{1}^{2}-L_{0}^{2}\right) \right] . \label{WDELTAE}
\end{equation}%
Comparing (\ref{WDELTAE}) with (\ref{DIFLC}) its clear that JE is verified, what was expected (see Ref. \cite{ENGEL}).

\section{Visibility of Interference Fringes and its connection with JE}

\label{Exp}
In this section, we verify the possibility of an experimental
realization procedure. We assume that the harmonic oscillator is a
microwave field stored in a high-$Q$ superconducting cavity.
The field state can be monitored by establishing an interaction
with a Rydberg atom \cite{cavidade,rydberg,livro}.
Rydberg atoms have suitable properties for use as probes of
even weak electromagnetic fields, such as high dipole moments, which ensure
high coupling strengths, and high mean lifetimes.
We consider a non-demolition measurement procedure \cite{nondemolation}
of the number of photons contained in the electromagnetic field by
setting a dispersive interaction between it and each Rydberg atom.
The number of photons in the cavity is probed by Ramsey interferometry \cite{ramsey},
and this measure allows obtaining some information about the field state inside
the cavity.

\begin{figure}[!h]
  \centering
  \includegraphics[scale=0.6,angle=00]{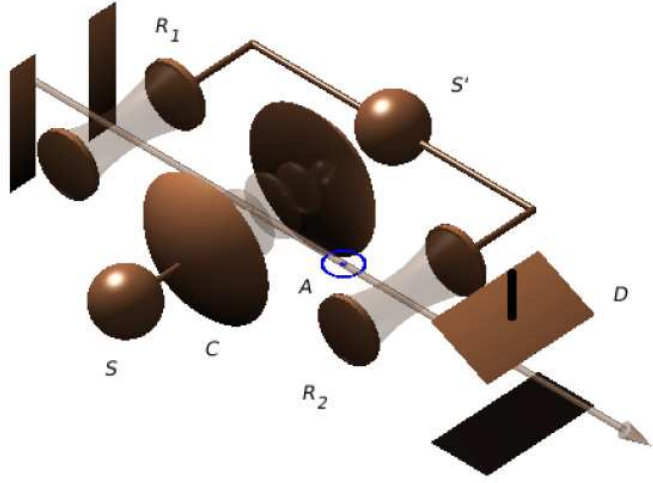}
  \caption{
  Schematic representation of the apparatus used in a typical
  Ramsey interferometry with Rydberg atoms. A Rydberg atom $A$, in general,
  an atom of an alkali element, is  prepared a highly excited electronic level
  $\left|g\right\rangle$ and it is sent through the apparatus.  The two Ramsey zones,
  $R_1$ and $R_2$, are low-$Q$ cavities devised to change the atomic states as
  $\left|g\right\rangle \rightarrow  \left(\left|g\right\rangle+\left|f\right\rangle\right)/\sqrt{2}$
  and $\left|f\right\rangle \rightarrow  \left(-\left|g\right\rangle+\left|f\right\rangle\right)/\sqrt{2}$.
  Despite the low  mean number of photons inside the two Ramsey zones,
  from the practical point of view, the atom sees a classical field there, so much
  that the atom leaves the Ramsey zones in a non-entangled state.
  In the superconducting microwave cavity $C$ the atom interacts dispersively with the cavity field.
  This interaction glues different phase shifts in each atomic state that depends on the number of
  photons of the cavity field. So, right after the atom leaves the superconducting
  cavity $C$, the global state of atom plus the field inside it remains \textit{entangled}.
  The detector $D$ measures the atom at  $\left|f\right\rangle$. Repeating the process
  under slight different conditions (for example, changing the frequency of the mode inside
  the Ramsey zones) a interferometric pattern is produced, and the information
  about the number of photons of the mode inside the cavity $C$ can be extracted.
   }
   \label{haroche}
\end{figure}

\begin{figure}[!h]
  \centering
  \includegraphics[scale=0.4,angle=-90]{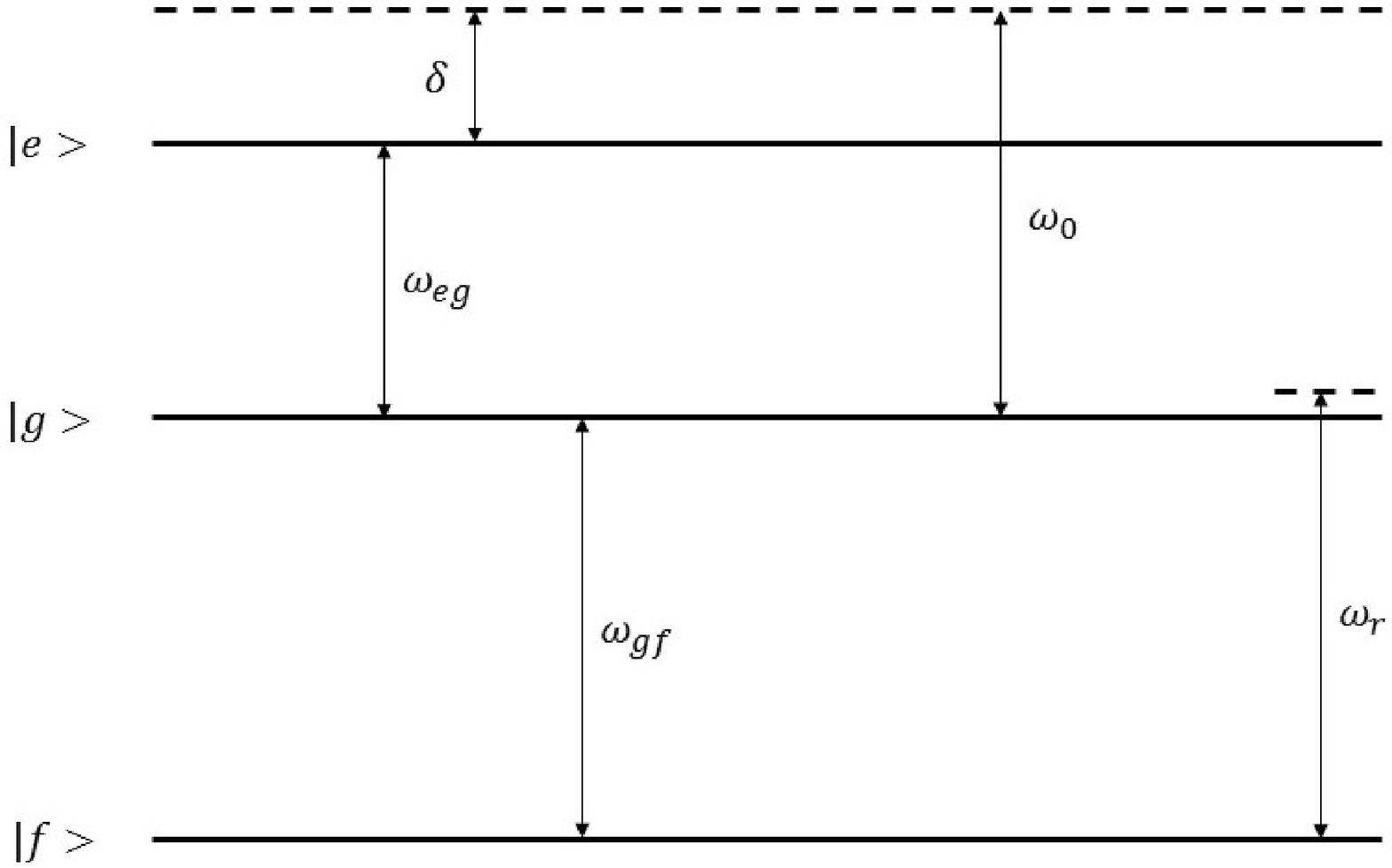}
  \caption{
 Three-level atom. The states $|e\rangle$ and $|f\rangle$ have the same parity and are opposed to the parity of  $|g\rangle$. The field in the superconducting cavity has frequency $\omega_0$ and is de-tuned of  $\delta=\omega_0
  -\omega_{eg}$ from transition frequency
$\omega_{eg}=(E_e-E_g)/\hbar$ between levels $|e\rangle$ and
  $|g\rangle$. The Ramsey zones \cite{ramsey} have frequencies $\omega_r$ and are close to the transition transition sintony frequency $\omega_{gf}=(E_g-E_f)/\hbar$ between levels $|g\rangle$ and $|f\rangle$.}\label{3niveis}
\end{figure}

A schematic representation of the experimental setup
is illustrated in FIG. \ref{haroche}.
We consider a three-level Rydberg atom, as illustrated in Fig.
\ref{3niveis}.
The three-level atom is sent through an apparatus as schematized in FIG. \ref{haroche}.
The atom, when passing through $ C $, will interact dispersively with the atom inside it and
the interest Hamiltonian is
\begin{equation}\label{h1}
\hat{H}=\hbar\omega_0\biggl(\hat{a}^\dag \hat{a}+\frac{1}{2}\biggr) +
E_e|e\rangle\langle e|+ E_g|g\rangle\langle g|+ E_f|f\rangle\langle
f| +\hbar\omega\Bigl[\bigl(\hat{a}^\dag \hat{a} +1\bigr)|e\rangle\langle
e|-\hat{a}^\dag \hat{a}|g\rangle\langle g|\Bigr],
\end{equation}
where $\hat{a}^\dag$ ($\hat{a}$) is the creation (annihilation) operator
acting on the field state inside the cavity $C$, $|i\rangle$ is the $i$-th
atomic level, defined as $i=e$, $g$ and $f$, $E_i$ is the
corresponding energy of the $i$th level and $\omega_0$ is the field
frequency in $C$, $\omega=\Omega_{0}^{2}/4\delta$ is the coupling
constant in the dispersive regime, $\Omega_{0}$ is the vacuum Rabi
frequency inside cavity $C$ and $\delta$ is the atom-field detuning
between transition frequency of the energy levels $|e\rangle$ and $|g\rangle$, $\omega_{eg}=(E_e-E_g)/\hbar$, and the frequency of the stored mode in $C$, $\omega_0$.

Without lost of generality, equation (\ref{h1}) can be presented as
\begin{equation}\label{h2}
\hat{H}=\hat{H}_0+\hat{H}_I,
\end{equation}
where
\begin{eqnarray}
\hat{H}_0&=&\hbar\omega_0\biggl(\hat{a}^\dag \hat{a}+\frac{1}{2}\biggr) + (E_e+\hbar\omega)|e\rangle\langle e|+ E_g|g\rangle\langle g|+ E_f|f\rangle\langle f|, \label{h0}\\
\hat{H}_I&=&\hbar\omega\,\hat{a}^\dag \hat{a}\,\bigl(|e\rangle\langle
e|-|g\rangle\langle g|\bigr). \label{hi}
\end{eqnarray}

We observe that $[\hat{H}_0,\hat{H}_I]=0$, then, in the interaction picture we
have an arbitrary state of the field in cavity $C$, it is given by

\begin{equation}\label{campo1}
\rho_{F}(0)=\sum_{i,j}\rho_{i,j}|i\rangle\langle j|,
\end{equation}
an atom is sent to interact with the field in cavity $C$, this atom
is previously prepared in the state
\begin{equation}\label{atomo1}
\rho_{A}(0)=g|g\rangle\langle g|+f|f\rangle\langle
f|+\bigl[\,x|g\rangle\langle f|+c.h.\bigr],
\end{equation}
with $g+f=1$ and $|x|^2\leq gf$. After a time interval $\Delta t$,
the atom-field state in the interaction picture, is given by
\begin{equation}\label{global1}
\rho(\Delta t)=e^{-iH_I\Delta
t/\hbar}\rho_{F}(0)\rho_{A}(0)e^{iH_I\Delta t/\hbar}
\end{equation}
taking the trace in field variables in time $\Delta t$ we obtain the
atomic state that is
\begin{equation}\label{atomo2}
\rho_{A}(\Delta t)= \mathrm{Tr}_{F}\bigl[\rho(\Delta t)\bigr].
\end{equation}
As the atom goes through $R_2$ the states $|g\rangle$ and
$|f\rangle$ will be entangled with a relative phase $\phi$, as we
can observe in FIG. \ref{haroche}. After that, the atom is
measured in $D$ and as we change $\phi$, the Ramsey interference
fringes appear. The visibility $\mathcal{V}$ of the interference
fringes pattern is proportional to the absolute value of coherence's
term of the state (\ref{atomo2}) and can be obtained by

\begin{equation}\label{visi1}
\mathcal{V}(\Delta t)=2\Bigl|\mathrm{Tr}\bigl[\rho_{A}(\Delta
t)|g\rangle\langle f|\bigr]\Bigr|,
\end{equation}
as we can see in reference \cite{bergou}.

The visibility $\mathcal{V}$ depends on field state
eq.(\ref{campo1}), then the interference fringes carries field state
information, it becomes clear as we analyze equation (\ref{atomo2})
in deeper, we have:
\begin{eqnarray}
\rho_{A}(\Delta t) &=& \sum_{m}\langle m|e^{-iH_I\Delta
t/\hbar}\rho_{F}(0)\rho_{A}(0)e^{iH_I\Delta t/\hbar}|m\rangle
\nonumber\\
                   &=& \sum_{m}\langle m|e^{-iH_I\Delta t/\hbar}\rho_{F}(0)e^{iH_I\Delta t/\hbar}e^{-iH_I\Delta
                   t/\hbar}\rho_{A}(0)e^{iH_I\Delta t/\hbar}|m\rangle \nonumber\\
                   &=& \sum_{m,n}\langle m|e^{-iH_I\Delta t/\hbar}\rho_{F}(0)e^{iH_I\Delta t/\hbar}|n\rangle\langle n|e^{-iH_I\Delta t/\hbar}\rho_{A}(0)e^{iH_I\Delta t/\hbar}|m\rangle \label{desen1}.
\end{eqnarray}
Taking (\ref{campo1}) into (\ref{desen1}), we obtain
\begin{eqnarray}
\rho_{A}(\Delta t) &=& \sum_{m,n} \rho_{m,n}e^{-i\omega \Delta
t(m-n)\bigl(|e\rangle\langle e|-|g\rangle\langle g|\bigr)}
                        \delta_{n,m}e^{-i\omega \Delta t(n-m)|e\rangle\langle e|}e^{i\omega \Delta tn|g\rangle\langle g|}\rho_{A}(0)
                        e^{-i\omega \Delta tm|g\rangle\langle g|}\nonumber\\
                   &=& \sum_{n,n} \rho_{n,n}e^{i\omega \Delta tn|g\rangle\langle g|}\rho_{A}(0)
                        e^{-i\omega \Delta tn|g\rangle\langle g|}\nonumber\\
                   &=& g|g\rangle\langle g|+f|f\rangle\langle f|+\bigl[\,x\sum_{n}\rho_{n,n}e^{in\omega \Delta t}
                        |g\rangle\langle f|+c.h.\bigr] \label{desen2}.
\end{eqnarray}
We observe that the term  $\sum_{n}\rho_{n,n}e^{in\omega \Delta t}$
in the equation (\ref{desen2}) can be written as
\begin{eqnarray}
  \sum_{n}\rho_{n,n}e^{in\omega \Delta t} &=&  \sum_{n}\langle n|\rho_{F}(0)e^{i\omega \Delta t a^\dag a}|n\rangle \nonumber\\
                                   &=& \mathrm{Tr}\Bigl[\rho_{F}(0)e^{i\omega \Delta t a^\dag a}\Bigr] \label{formatraco}.
\end{eqnarray}
Now, introducing (\ref{formatraco}) into \ref{desen2}) the field
state is given by
\begin{equation}\label{atomo3}
\rho_{A}(0)=g|g\rangle\langle g|+f|f\rangle\langle
f|+\bigl[\,\bar{x}|g\rangle\langle f|+c.h.\bigr],
\end{equation}
where $\bar{x}=x\,\mathrm{Tr}\Bigl[\rho_{F}(0)e^{i\omega \Delta t
a^\dag a}\Bigr]$. We can obtain the Ramsey interference fringes
visibility, equation (\ref{visi1}) for the atom in the
state(\ref{atomo3}), then
\begin{equation}\label{visi2}
\mathcal{V}(\Delta
t)=\mathcal{V}_{0}\biggl|\mathrm{Tr}\Bigl[\rho_{F}(0)e^{i\omega
\Delta t a^\dag a}\Bigr]\biggr|.
\end{equation}
The visibility clearly depends on field initial  state. In equantion
(\ref{visi2}) we defined $\mathcal{V}_{0}=2|x|$, it is the
visibility of the vacuum state. Form now on, we consider the optimum
case where$\mathcal{V}_{0}=1$. This occurs when $|x|=1/2$ and the
atom state is $(|g\rangle+e^{i\theta}|f\rangle)/\sqrt{2}$, and
$\theta$ is a arbitrary relative phase.

\subsection{The field state in a equilibrium Thermal State}

Following the previous mathematical procedure, easily we can obtain
the visibility when we have in $C$ a field initially in a thermal
given by
\begin{equation}\label{termico}
\rho_{Th}=\frac{e^{-\beta\bar{H}}}{Z},
\end{equation}
where $\bar{H}=\hbar\omega_0\bigl(a^\dag a+1/2\bigr)$ is the field
Hamiltonian that is in thermal equilibrium and
$Z=\mathrm{Tr}\bigl(e^{-\beta\bar{H}}\bigr)$ , as usual, is the
partition function. Then, after some algebra we obtain
\begin{equation}\label{visitemico1}
\mathcal{V}_{Th}(\Delta
t)=\dfrac{\sinh(\beta\hbar\omega_{0}/2)}{\sqrt{\sinh^2(\beta\hbar\omega_{0}/2)
+\sin^2(\omega \Delta t/2)}}.
\end{equation}
For practical proposes, the perfect choice of the interaction time
can be obtained if we adopt $\omega\Delta t=\pi$, then
equation(\ref{visitemico1}) can now be written as
\begin{equation}\label{visitermico2}
\mathcal{V}_{Th}(\pi/\omega)=\tanh(\beta\hbar\omega_{0}/2).
\end{equation}
It is interesting to note that equation (\ref{visitemico1}) can be
simplified in two limiting cases
\begin{equation}\label{limites1}
\mathcal{V}_{Th}(\pi/\omega)\approx\left\{\begin{array}{rcc}
                                      \beta\hbar\omega_{0}/2, & \mathrm{se} & \beta\hbar\omega_{0} \ll 1, \\
                                        & & \\
                                      1, & \mathrm{se} & \beta\hbar\omega_{0}\gg 1.
                                    \end{array}
                                    \right.
\end{equation}

\subsection{The field state in a Displaced Thermal State}

The Hamiltonian for a displaced field in a cavity $C$ is given by
\begin{equation}\label{hdeslocado1}
\tilde{H}=\hbar\omega_{0}\Bigl(a^\dag a +\frac{1}{2}\Bigr)
+\hbar\omega_0\bigl(\alpha a^\dag+\alpha^{*} a\bigr).
\end{equation}
where  $\alpha$ is the displacement magnitude, observe that it is directly related with the protocol term $\tilde{L}$ in equation (\ref{HAMOP22}). The Hamiltonian
(\ref{hdeslocado1}) can be written as
\begin{equation}\label{hdeslocado2}
\tilde{H}=D^{\dag}(\alpha)\bar{H}D(\alpha)-\hbar\omega_0|\alpha|^2,
\end{equation}
where $D$ is the displacement operator defined as
$D(\alpha)=\exp\bigl(\alpha a^\dag-\alpha^{*} a\bigr)$. In this
case, the thermal equilibrium state is
\begin{equation}\label{termicodesl}
\rho^{(\alpha)}_{Th}=\frac{e^{-\beta\tilde{H}}}{Z^{(\alpha)}},
\end{equation}
now the partition function is
$Z^{(\alpha)}=\mathrm{Tr}\bigl(e^{-\beta\tilde{H}}\bigr)$. Easily we
can show that
\begin{eqnarray}
\rho^{(\alpha)}_{Th} &=& D^{\dag}(\alpha)\rho_{Th}D(\alpha), \label{termicodesl2} \\
  Z^{(\alpha)} &=& Z\,e^{\beta\hbar\omega_0|\alpha|^2}. \label{particao}
\end{eqnarray}
Now we can obtain the new visibility in the same way we obtained for
the Thermal State (previous subsection), but now we consider the
state (\ref{termicodesl}). Just replacing  $\rho_{F}(0)$ with
$\rho^{(\alpha)}_{Th}$ in eq.(\ref{visi2}), then, after some algebra
we find the visibility as

\begin{equation}\label{visideslocado1}
\mathcal{V}^{(\alpha)}_{Th}(\Delta t)=\mathcal{V}_{Th}(\Delta
t)\exp\Bigl\{-2|\alpha|^2\frac{\sin^{2}(\omega\Delta t/2)
\sinh(\beta\hbar\omega_0/2)\cosh(\beta\hbar\omega_0/2)}{\sinh^2(\beta\hbar\omega_0/2)+\sin^{2}(\omega\Delta
t/2)}\Bigr\}.
\end{equation}
Again, assuming a interaction time as $\Delta t=\pi/\omega$, then
the equation (\ref{visideslocado1}) can be simplified as
\begin{equation}\label{visideslocado2}
\mathcal{V}^{(\alpha)}_{Th}(\pi/\omega)=\mathcal{V}_{Th}(\pi/\omega)\,e^{-2|\alpha|^2
\tanh(\beta\hbar\omega_0/2)}.
\end{equation}
Now we will investigate the limiting cases, as in the previous
subsection. The main difference is that the displacement magnitude
plays an important role. We have now four situations, they are
\begin{enumerate}
  \item For $|\alpha|\sim 1$ we have:
        \begin{equation*}
        \frac{\mathcal{V}^{(\alpha)}_{Th}(\pi/\omega)}{\mathcal{V}_{Th}(\pi/\omega)}
        =\left\{\begin{array}{rcc}
        1-|\alpha|^2\beta\hbar\omega_{0}, & \mathrm{if} & \beta\hbar\omega_{0} \ll 1, \\
        & & \\
        e^{-2|\alpha|^2}, & \mathrm{if} & \beta\hbar\omega_{0}\gg 1.
        \end{array}
        \right.
        \end{equation*}
  \item For $|\alpha|\gg 1$ we have:
        \begin{equation*}
        \frac{\mathcal{V}^{(\alpha)}_{Th}(\pi/\omega)}{\mathcal{V}_{Th}(\pi/\omega)}
        =\left\{\begin{array}{rcc}
        e^{-|\alpha|^2\beta\hbar\omega_{0}}, & \mathrm{if} & \beta\hbar\omega_{0} \ll 1, \\
        & & \\
        e^{-2|\alpha|^2}, & \mathrm{se} & \beta\hbar\omega_{0}\gg 1.
        \end{array}
        \right.
        \end{equation*}
\end{enumerate}
This can be summarized by
\begin{equation}\label{limitesd}
\frac{\mathcal{V}^{(\alpha)}_{Th}(\pi/\omega)}{\mathcal{V}_{Th}(\pi/\omega)}
=\left\{\begin{array}{rcc}
        e^{-|\alpha|^2\beta\hbar\omega_{0}}, & \mathrm{if} & \beta\hbar\omega_{0}/2 \ll 1 \\
        & & \\
        e^{-2|\alpha|^2}, & \mathrm{if} & \beta\hbar\omega_{0}/2\gg 1
        \end{array}
\right.
\end{equation}
and this is the limiting case form the displaced Thermal State.

\subsection{Mathematical relationship between the variation in
Helmholtz free energy and the variation of visibility}

The Helmholtz free energy can be defined as
\begin{equation}\label{helmholtz}
F=-\frac{1}{\beta}\ln(Z_0)
\end{equation}
where $Z_0$ is the state partition function. The variation in
Helmholtz free energy $\Delta F$ from the initial state given
(\ref{termico} to the final displaced state given by
(\ref{termicodesl}), can be obtained by the respective partition
functions, given by
\begin{eqnarray*}
Z&=&\mathrm{Tr}\bigl(e^{-\beta\bar{H}}\bigr)\\
Z^{(\alpha)}&=&Z\,e^{\beta\hbar\omega_0|\alpha|^2},
\end{eqnarray*}
then we obtain
\begin{equation}\label{varH1}
\Delta F=-\hbar\omega_0|\alpha|^2.
\end{equation}
Now, the term  $e^{-\beta\Delta F}$ can be represented as
\begin{equation}\label{jarzy1}
e^{-\beta\Delta F}=e^{|\alpha|^2\beta\hbar\omega_0}.
\end{equation}
From the equation (\ref{visitemico1}) we can write
\begin{equation}\label{sin}
\sin^2(\omega \Delta t/2)=\dfrac{1-\bigl[\mathcal{V}_{Th}(\Delta
t)\bigr]^2}{\bigl[\mathcal{V}_{Th}(\Delta
t)\bigr]^2}\sinh^2(\beta\hbar\omega_{0}/2).
\end{equation}
Substituting (\ref{sin}) into (\ref{visideslocado1}) we get the relation
%
\begin{equation}\label{relacao}
\Biggl[\dfrac{\mathcal{V}_{Th}}
{\mathcal{V}_{Th}^{(\alpha)}}\Biggr]^{1/(1-\mathcal{V}_{Th}^{2})}=e^{|\alpha|^2\sinh(\beta\hbar\omega_{0})}
\end{equation}
%
between the visibilities independent of the interaction time $\Delta t$.
In case that $\beta\hbar\omega_{0} \ll 1$ it is immediate that
\begin{equation}\label{jarzyvisi}
e^{-\beta\Delta
F}=\Biggl[\dfrac{\mathcal{V}_{Th}}
{\mathcal{V}_{Th}^{(\alpha)}}\Biggr]^{1/(1-\mathcal{V}_{Th}^{2})}.
\end{equation}
This result shows that the variation in Helmholtz free energy, for
this experimental set-up, can be obtained in terms of visibility, it
means that the term  $e^{-\beta\Delta F}$ from Jarzynski
\cite{jarzy} equality can be experimentally obtained from visibility
measurements.
\subsection{Quantum work measurement}


There are many possibilities for choosing the quantum work operator
\cite{Deffner, Valente, Talkner2}, and the best definition still an
open question \cite{Dorner2013,Mazzola2013,Deffner, Valente,
Talkner2}. If we consider the definition $\Delta E$ then the JE is
validly and we can use the visibility to estimate the work done on
the cavity, in this case we have

\begin{equation}\label{visiwork}
e^{-\beta\Delta F}=\Biggl[\dfrac{\mathcal{V}_{Th}}
{\mathcal{V}_{Th}^{(\alpha)}}\Biggr]^{1/(1-\mathcal{V}_{Th}^{2})}=\langle
e^{-\beta \Delta W}\rangle.
\end{equation}
%
The Cavity quality factor plays an important role on this case,
since the calculations where carried out without considering
dissipation. That means that the protocol time should be small
comparing with the life time \cite{Brune2008} of the field in the
cavity.

Another  quantum work definition is constructed in \cite{Talkner2},
and it gives an different result from $\Delta E$, they demonstrate
that quantum correlation function
\begin{equation}\label{correlacao}
G(u)=\mathrm{Tr}\Bigl[U^{\dag}(t)e^{i u
H(t)}U(t)e^{-iuH(0)}\rho(0)\Bigr],
\end{equation}
%
with $U(t)$ solution of the equation $i\hbar \partial U(t)/\partial
t=H(t)U(t)$ and $U(0)=\mathbf{1}$, when $u=i\beta$ contains all
available statistical information about the work such as the
averaged exponentiated work $\langle \exp(-\beta W) \rangle$. When
we calculate the eq.(\ref{correlacao}) for our experimental
proposal, again we find precisely that $ e^{-\beta\Delta F}=\langle
e^{-\beta W} \rangle$, again the visibility is a useful tool to determine the free energy variation and consequently the averaged exponential work.

%

%

\section{Conclusions}
\label{Conclusions} We have shown that Jarzynski's theorem can be
tested experimentally in the context of superconducting cavities. In
particular, we find a direct relationship between the visibility and
the free energy variation. Considering the work operator $\Delta E$
we have shown that it is possible to use de Jarzynzki equality to
determine the work done or extracted at a superconducting cavity by
measuring the fringes visibility, that is commonly used in cavity
experiments. In terms of JE, we can also obtain the state of field in the cavity with the visibility measure, since we know the initial state. Taking into account that visibility measurement is simpler than usual state measurements, this approach is very efficient.

\section*{Acknowledgements}
RCF and ACO  gratefully acknowledge the support of Brazilian agency Funda\c{c}\~{a}o de Amparo a Pesquisa do Estado de Minas Gerais
(FAPEMIG) through grant No. APQ-01366-16.


\begin{thebibliography}{99}
\bibitem{jarzy} C. Jarzynski, {\it Phys. Rev. Lett.} {\bf 78}, 2690 (1997).

\bibitem{JAR1} C. Jarzynski, Eur. Phys. J B 64, 331 (2008).
\bibitem{JAR2} C. Jarzynski, Annu. Rev. Condens. Matter Phys. 2, 329 (2011).
\bibitem{JAR3} E. Boksenbojm, B. Wynants and C. Jarzynski, Physica A 389,
4406 (2010).
\bibitem{Crooks} G. E. Crooks, J. Stat. Mech.: Theor. Exp., P10023 (2008).


\bibitem{Bochkov} G. N. Bochkov and Y. E. Kuzovlev, Sov. Phys. JETP, 45, 125 (1977)


\bibitem{Morgado} W.A.M. Morgado and D.O. Soares-Pinto, Phys. Rev. E 82,
021112 (2010).

\bibitem{EXJAR1} Shoichi Toyabe,    Takahiro Sagawa,    Masahito Ueda,  Eiro Muneyuki   and Masaki Sano, Nature Physics 6, 988?992 (2010) doi:10.1038/nphys1821


\bibitem{Minh} D.D.L. Minh and A.B. Adib, Phys. Rev. E 79, 021122 (2009).


\bibitem{Liphardt} Liphardt J, Dumont S, Smith S B, Tinoco I Jr and
Bustamante C 2002 Equilibrium information from nonequilibrium
measurements in an experimental test of Jarzynskis equality Science
296 1832 5.

\bibitem{EXJAR3} F. Douarche, S. Ciliberto, A. Petrosyan and I. Rabbiosi, EPL, 70, Number 5 (2005).

\bibitem{Hoang} Hoang, Thai M. and Pan, Rui and Ahn, Jonghoon and Bang, Jaehoon and Quan, H. T. and Li, Tongcang, Experimental Test of the Differential Fluctuation Theorem and a Generalized Jarzynski Equality for Arbitrary Initial States,  Phys. Rev. Lett. 120, 080602 (2018).

\bibitem{Hummer} Hummer G, Szabo A. Free-energy reconstruction from nonequilibrium single molecule experiments. Proc Natl Acad Sci USA.  v 98, p 3658-3661, (2001).

\bibitem{Yukawa} S. Yukawa, J. Phys. Soc. Jpn. 69, 2367, (2000)

\bibitem{Allahverdyan} A. E. Allahverdyan and T. M. Nieuwenhuizen, Phys. Rev. E 71, 066102 (2005)

\bibitem{ENGEL} A. Engel and R. Nolte, EPL 79, 10003 (2007).

\bibitem{Gelin} M. F. Gelin and D. S. Kosov, Phys. Rev. E, 78 011116 (2008).

\bibitem{Talkner} P. Talkner, E. Lutz and P. H\"anggi, Phys. Rev. E 75,
050102(R) (2007).

\bibitem{Talkner1} P. Talkner and P. H\"anggi, J. Phys. A: Math. Theor. 40, F569 (2007).


\bibitem{Talkner2} P. Talkner and P. H\"anggi, Phys. Rev. E 93,
022131 (2016).

\bibitem{Talkner3} P.H\"anggi and P. Talkner, Nat. Phys., 11, 108
(2015).

\bibitem{Talkner4} P. Talkner, P.H\"anggi and and M. Morillo, Phys. Rev. E,  77, 051131
(2008).

\bibitem{Talkner5} P. Talkner, M. Campisi and P. H\"anggi, J. Stat. Mech: Theor. Exp. P02025, (2009).

\bibitem{Campisi} M. Campisi, P. Talkner and

\bibitem{Kurchan} J. Kurchan e-print arXiv:cond-mat/0007360 (2000).

\bibitem{Tasaki} H. Tasaki e-print arXiv:cond-mat/0009244 (2000).



\bibitem{Esposito} M. Esposito and S. Mukamel, Phys. Rev. E, 73, 046129 (2006).


\bibitem{Paz} Augusto J. Roncaglia and Federico Cerisola and Juan Pablo Paz, Phys. Rev.
Lett. 113, 250601 (2014).


\bibitem{EXJAR2} Shuoming An,   Jing-Ning Zhang,    Mark Um,    Dingshun Lv,Yao Lu, Junhua Zhang,   Zhang-Qi Yin,   H. T. Quan  and Kihwan Kim Nature Physics 11, 193?199 (2015) doi:10.1038/nphys3197

\bibitem{EXJAR4} Nolan C. Harris, Yang Song, and Ching-Hwa Kiang
Phys. Rev. Lett. 99, 068101 (2007).


\bibitem{cavidade} J. M. Raimond, M. Brune, \& S. Haroche, {\it Rev. Mod. Phys.} {\bf 73}, 565 (2001).


\bibitem{Nogues} G. Nogues, A. Rauschenbeutel, S. Osnaghi, P. Bertet, M. Brune, J.
M. Raimond, S. Haroche, L. G. Lutterbach, and L. Davidovich Phys.
Rev. A 62, 054101 (2000).


\bibitem{Davidovich2016} L Davidovich,  Physica Scripta, {\bf 91}, 6 (2016).


\bibitem{Haroche1991} S. Haroche, M. Brune and J. M. Raimond, EPL (Europhysics Letters),
{\bf 14},  1, 19 (1991).

\bibitem{rydberg} R. G. Hulet, \& D. Kleppner, {\it Phys. Rev. Lett.} {\bf 51}, 1430
(1983);\\P. Nussenzveig, F. Bernardot, M. Brune, J. Hare, J. M.
Raimond, S. Haroche \& W. Gawlik, {\it Phys. Rev. A} {\bf 48}, 3991
(1993).

\bibitem{Faria1999} J. G. Peixoto de Faria and M. C. Nemes Phys. Rev. A 59,
3918 (1999).

\bibitem{Oliveira2003} Oliveira, A. C. and Nemes,M.C. and Romero,K.M.Fonseca, %
\newblock Phys Rev. E \textbf{68}, 036214 (2003).









 \bibitem{Lemos2018} Humberto C.F.Lemos, Alexandre C.L.Almeida, Barbara Amaral, Adelcio C.Oliveira, Physics Letters A , 382 (2018), 823-836.



\bibitem{Cerisola} Cerisola, F. and Margalit, Y. and Machluf, S. and Roncaglia, A. J. and Paz, J. P.,  Folman, R., Using a quantum work meter to test non-equilibrium fluctuation theorems, Nature Communications, 1241, v. 8 (2017).

\bibitem{Hijar} H\'ijar, H. and de Z\'arate, J. M. O. Jarzynski's equality illustrated by simple examples, European Journal of Physics, (2010), 31,
1097.


\bibitem{livro} S. Haroche, \& J. M. Raimond, {\it Exploring the Quantum: Atoms, Cavities and Photons} (Oxford Univ. Press, New York, 2006).

\bibitem{nondemolation} M. Brune, S. Haroche, J. M. Raimond, L. Davidovich, \& N. Zagury, {\it Phys. Rev. A} {\bf 45}, 5193 (1992).

\bibitem{ramsey} N. F. Ramsey, {\it Molecular Beams} (Oxford Univ. Press, Oxford, 1985);\\ J. I. Kim, K. M. Fonseca Romero, A. M. Horiguti, L. Davidovich, M. C. Nemes, \& A. F. R. de Toledo Piza, {\it Phys. Rev. Lett.} {\bf 82}, 4737 (1999).

\bibitem{bergou} M. Jakob, and J. Bergou, {\it Opt. Commun}, {\bf 283}, 827(2010).\\
                 M. Jakob, and J. Bergou, e--print arXiv:quant-ph/0302075 (2003).



\bibitem{Mondaini} Felipe Mondaini and L. Moriconi, Physics Letters A
378, 1767-1772 (2014).

\bibitem{Brune2008} M. Brune, J. Bernu, Guerlin, S. Deléglise, Sayrin, S. Gleyzes, S. Kuhr,I. Dotsenko,
J. M. Raimond, and S. Haroche, Phys. Rev. Lett. 101, 240402 (2008).

\bibitem{Ngo2016} V. A. Ngo, I. Kim,  T. W. Allen, and S. Y. Noskov, J. Chem. Theory Comput. 12, 1000 (2016).

\bibitem{Deffner} S. Deffner, J. P. Paz, and W. H. Zurek, PHYSICAL REVIEW E 94, 010103(R) (2016).




\bibitem{Mazzola2013} L. Mazzola, G. De Chiara, and M. Paternostro, Phys. Rev. Lett. 110, 230602
(2013).

\bibitem{Collin2005} D. Collin, F. Ritort, C. Jarzynski, S. B. Smith, I. Tinoco, and
C. Bustamante, Nature (London) 437, 231 (2005).

\bibitem{Dorner2013} R. Dorner, S. R. Clark, L. Heaney, R. Fazio, J. Goold, and V. Vedral,Phys. Rev. Lett. 110, 230601
(2013).

\bibitem{Valente} D. Valente, F. Brito, R. Ferreira, T. Werlang,
arXiv:1709.09677v1 (2017).























\end{thebibliography}
\end{document}